\documentclass[submission, Phys]{SciPost}
\newcommand{\alexket}[1]{\left|#1\right\rangle}
\newcommand{\alexbra}[1]{\left\langle#1\right|}

\begin{document}

% TODO: write your article's title here.
% The article title is centered, Large boldface, and should fit in two lines
\begin{center}{\Large \textbf{
A short story of quantum and information thermodynamics
}}\end{center}

% TODO: write the author list here. Use initials + surname format.
% Separate subsequent authors by a comma, omit comma at the end of the list.
% Mark the corresponding author with a superscript *.
\begin{center}
A. Auff\`eves\textsuperscript{1*}
\end{center}

% TODO: write all affiliations here.
% Format: institute, city, country
\begin{center}
{\bf 1} Universit\'e Grenoble Alpes, CNRS, Grenoble INP, Institut N\'eel, 38000 Grenoble, France\\
% TODO: provide email address of corresponding author
* alexia.auffeves@neel.cnrs.fr
\end{center}

\begin{center}
\today
\end{center}

% For convenience during refereeing: line numbers
%\linenumbers

\section*{Abstract}
{\bf
% TODO: write your abstract here.
This Colloquium is a fast journey through the build-up of key thermodynamical concepts, i.e. work, heat and irreversibility -- and how they relate to information. Born at the time of industrial revolution to optimize the exploitation of thermal resources, these concepts have been adapted to small systems where thermal fluctuations are predominant. Extending the framework to quantum fluctuations is a great challenge of quantum thermodynamics, that opens exciting research lines e.g. measurement fueled engines or thermodynamics of driven-dissipative systems. On a more applied side, it provides the tools to optimize the energetic consumption of future quantum computers.}

% TODO: include a table of contents (optional)
% Guideline: if your paper is longer that 6 pages, include a TOC
% To remove the TOC, simply cut the following block
\vspace{10pt}
\noindent\rule{\textwidth}{1pt}
\tableofcontents\thispagestyle{fancy}
\noindent\rule{\textwidth}{1pt}
\vspace{10pt}

\section{Legacy of classical thermodynamics}

\vspace{0.5cm}

\subsection{Macroscopic thermodynamics}
Thermodynamics was developed in the XIXth century, providing a unified framework between mechanical sciences and thermometry. At the time, the motivation was very practical, namely use temperature to put bodies into motion - as clearly indicated by its name. In other words, the goal was to design and optimize thermal engines, i.e. devices that exploit the transformations of some ``working substance" to convert heat into work. Work and heat are two ways to exchange energy, and according to the first law of thermodynamics, it is possible to convert one into another. 

However, turning heat into work is like turning lead into gold: It has severe constraints. The most famous is Kelvin no-go statement: It is not possible to extract work cyclically from a single hot bath. This no-go statement turned out to be one of the expressions of the second law of thermodynamics, which deals with (ir)reversibility. This is how an initially applied area of physics turned out to deliver fundamental concepts like entropy and time arrow.

As a matter of fact, the first boundary between work and heat was intimately related to the (ir)reversible nature of their exchanges. The concept of work comes from mechanical sciences, and represents a form of energy that can be exchanged reversibly: In principle, there is no time arrow associated with work exchanges - at least those associated to conservative forces. Conversely, the heat exchanges between a body and thermal baths are in general not reversible: heat spontaneously flows from hot to cold bodies. In particular, if a body cyclically exchanges an amount of heat $Q$ with a hot bath of temperature $T_h$ and $-Q$ with a cold bath of temperature $T_c$, the irreversible nature of heat transfers is captured by  the phenomenological formula $Q(1/T_c - 1/T_h)\geq 0$, with equality if $T_c = T_h$. 

This observation led to define the entropy change of a body in contact with a bath at temperature $T$ as $\Delta S = Q_\mathrm{rev}/T$, where $Q_\mathrm{rev}$ is the amount of heat reversibly exchanged. More generally, any isothermal heat exchange follows the Clausius inequality $\Delta S - Q/T = \Delta_\mathrm{i} S \geq 0$. $ \Delta_\mathrm{i} S$ is the so-called entropy production that quantifies the irreversibility of the transformation. Introducing the system's internal energy $U$, and its free energy $F=U-TS$, Clausius inequality becomes 
\begin{equation} \label{F}
W - \Delta F = T \Delta_\mathrm{i} S \geq 0. 
\end{equation}
The meaning of Eq.\ref{F} is transparent: It is not possible to extract more work than the free energy of the system. Reciprocally, to increase the free energy of a system, one has to pay at least the same amount of work. 
Since they are natural consequences of the thermodynamic arrow of time, these inequalities are called fundamental bounds. Extending these bounds to the quantum realm is an important motivation for developing a quantum thermodynamics. 

Eq.\ref{F} provides intuitions on Kelvin no-go statement, as exemplified by the Carnot engine. In this paradigmatic device, work is extracted during the expansion of the gas while being coupled to the hot bath - increasing its entropy and thus lowering its free energy by an amount $T_h \Delta S$. $\Delta S$ is fixed by the engine settings (number of gas particles, minimal and maximal volume of the chamber). Therefore once the maximal volume has been reached, it is not possible to extract work anymore and one has to ``reset" the engine, i.e. bring it back to its initial settings by compressing the gas. If the compression is performed at the same temperature, no net work is extracted from the cycle, meaning that two baths at two different temperatures are required for work extraction.

\subsection{Information thermodynamics}
Latter Maxwell suggested that information could be used to sort out the molecules of the gas and lower its entropy, apparently at no work cost. A so called Maxwell demon applying such mechanism would obviously violates the Second Law and Kelvin no go, since no cold bath here would be needed to reset the engine - compression being realized for free, only using information.

It took one century to exorcize Maxwell's demon paradox. With the rise of information theory after the Second World War, it became clear that information was not some immaterial concept that could escape the laws of thermodynamics. This idea is captured by the famous "Information is physical" attributed to Landauer \cite{Landauer61}, one of the fathers of information thermodynamics (see also \cite{Bennett}, \cite{Jaynes}...). To understand how information and thermodynamics are related, Carnot engines are enlightening. However, instead of considering a gas made of a large number of particles, one can consider a ``single-particle gas" rather, that is either positioned on the left or on the right of the chamber. Left or right can be used to encode one bit of information. Denoting by $p$ the probability that the particle is on the left, the Shannon entropy of the probability distribution reads (in bits) $H[p] = - p \log_2(p) - (1-p) \log_2 (1-p)$. Now imagine that we know the particle is on the left. While it expands to eventually fill the whole volume, one bit of information is lost, such that the Shannon entropy change reads $\Delta H = 1$ bit. Conversely, from this elementary amount of information, one can extract some work. In agreement with the Second Law, the amount of extractable work is bounded by $W_0 = k_\mathrm{B} T \log 2$, where $T$ is the temperature of the chamber in which the expansion takes place. Here $k_\mathrm{B}$ stands for the Boltzmann constant. This is the basic principle of the so-called Szilard engine, that evidences the conversion of information into work.  

This conversion is reversible and its reverse has even stronger practical implications. Indeed, starting from an initial configuration where the particle has equal chances to be on the left or on the right, and then compressing it, e.g. to the left of the chamber, is what is called a RESET operation in information theory: Whatever the initial state of the bit, it ends up in the state ``0".  This operation is {\it logically} irreversible: When it is performed, the initial state cannot be traced back. However, it is extremely useful since initializing bits is the beginning of any computation. Formula \ref{F} evidences that resetting a bit has a work cost, that cannot be lower than $W_0 = k_\mathrm{B} T \log 2$ - the bound being reached when the operation is {\it thermodynamically} reversible: This is the famous Landauer's erasure work \cite{Landauer61}.
 
Of course, the single particle model is a convenient approach to quickly get an intuition of the main equations, but it is idealized. First experimental evidences of information to energy conversions (Szilard engines, Landauer's erasure) have been obtained around 2010-2012 \cite{Toyabe2010,Beirut2012}.

\section{Stochastic thermodynamics}
While introducing information thermodynamics, one has departed from the usual scenery of macroscopic thermodynamics that involved large amounts of particles as working substances. Within information thermodynamics, the working substance is now elementary since it solely involves one particle whose phase space reduces to two micro-states, 0 and 1. This new scenery is the one of ``stochastic thermodynamics", that deals with small enough working substances for fluctuations to become predominant  \cite{Sekimoto, Seifert2008}. 

In this new realm, the dynamics of the system results from the action of some external operator that drives the system to implement some protocol. The system's evolution is perturbed by a thermal bath that induces random, ``stochastic" fluctuations. Thus, the dynamics of the system is described by Markovian, stochastic trajectories is its phase space - one trajectory consisting in continuous sequences where the drive controls the system, intertwined by stochastic jumps imposed by the bath. 

This new realm sheds new light on the First Law. Work now corresponds to the part of energy exchanged with the controller during the continuous sequences. On the other hand, heat is defined as the part of energy stochastically exchanged during the jumps induced by the bath. From an energetic point of view, it appears that the heat/work boundary now reflects the boundary between noise and control. From that perspective, an engine is a device made to extract energy from noise, by rectifying the fluctuations it induces. 

The framework of stochastic thermodynamics also invites to reconsider the meaning of the Second Law. As a matter of fact, the laws of physics at the level of single particles are expected to be reversible, so where does irreversibility come from? There is a simple, operational answer to this question. Let us suppose that the system is initially prepared in a well-defined micro-state. The controller now implements a protocol aimed to bring the system into another micro-state. In the absence of a bath, the trajectory is perfectly deterministic. The controller is thus able to reverse the protocol, to bring back the system to its initial micro-state. However, if a bath remains coupled to the system during the protocol, the random perturbations it induces prevent the controller from perfectly reversing the trajectory followed by the system, making the protocol irreversible.

\subsection{Fluctuation theorems}
Interestingly, stochastic thermodynamics allows us to quantify the amount of irreversibility {\it per trajectory} $\gamma$. This is captured by the so-called stochastic entropy production $\Delta_\mathrm{i}S[\gamma]$, that is defined as \cite{Sekimoto, Seifert2008}

\begin{equation} \label{Si}
\Delta_\mathrm{i}S[\gamma] = \log \left( \frac{P_\mathrm{F}[\gamma]}{P_\mathrm{B}[\gamma^*]} \right).
\end{equation}
We have introduced the time-reversed trajectory $\gamma^*$, and the probability $P_\mathrm{F}[\gamma]$ (resp. $P_\mathrm{B}[\gamma^*]$) of the trajectory $\gamma$ (resp. $\gamma^*$) while the protocol is run forward (resp. backward). With this definition, $\Delta_\mathrm{i} S$ has no dimension. The meaning of Eq.\ref{Si} is obvious. Entropy production is positive if the forward trajectory $\gamma$ is more probable than its corresponding backward trajectory $\tilde{\gamma}$. 

Some particular trajectories may lead to a negative entropy production. However, this is not contradictory with the Second Law which deals with average values. Namely, the quantity that should remain positive is the average value of the entropy production given by $\langle \Delta_\mathrm{i}S[\gamma] \rangle_\gamma = \Sigma_{\gamma} P_\mathrm{F}[\gamma] \Delta_\mathrm{i}S[\gamma]$. This condition is automatically fulfilled. This is obvious by noticing that $ \langle \exp(-\Delta_\mathrm{i}S[\gamma] ) \rangle_\gamma= \Sigma_{\gamma} P_\mathrm{B}[\gamma]$. In most situations (at the noticeable exception of so-called {\it absolute irreversibility} \cite{Funo_2015}), this boils down to 

\begin{equation} \label{FT}
 \langle \exp(-\Delta_\mathrm{i}S[\gamma] ) \rangle_\gamma = 1,
\end{equation}
which is called the Integral Fluctuation Theorem (IFT). From the convexity of the exponential function, one easily gets that $ \langle \Delta_\mathrm{i}S[\gamma]  \rangle_\gamma \geq 0$, in agreement with the Second Law. The bound is saturated, if and only if $\Delta_\mathrm{i}S[\gamma] =0$ for all $\gamma$. This strong condition defines the equilibrium distribution. As we show below, it is fully equivalent to define the equilibrium distribution by imposing the micro-reversibility condition.  

From Eq.\ref{Si}, Eq.\ref{FT} is a tautology - but it is also the seed of so-called ``fluctuation theorems", Jarzynski Equality (JE) probably being the most famous \cite{Jarzynski97}. To recover it, one considers a system initially prepared at thermal equilibrium, then driven out of equilibrium by some external operator. It can be shown that the expression of the stochastic entropy production exactly matches the classical one, namely $\Delta_\mathrm{i}S[\gamma] = (W[\gamma] - \Delta F)/k_\mathrm{B}T$ (See below for a simple approach). The IFT reads $\langle \exp (- W[\gamma] / k_\mathrm{B}T) \rangle_\gamma = \exp (- \Delta F / k_\mathrm{B}T) $. This is JE, which has been experimentally verified on many different platforms (See, e.g. \cite{Douarche, Bustamante, Beirut2012}). 

\subsection{Two-points trajectories}
While simple, two-points trajectories are interesting since they allow us to develop many useful intuitions. Let us consider a system with micro-states denoted by $\sigma_i$, $i$ being an integer. The stochastic behavior of the system is fully captured by the time-dependent probability of jumping from the state $j$ to the state $i$. Here we take this probability as a constant for the sake of simplicity and denote it $P[\sigma_i|\sigma_j]$. Denoting as $\gamma$ an elementary trajectory $\gamma = (\sigma_i, \sigma_j)$, its forward (resp. backward) probability reads $P_\mathrm{F}[\gamma] = p_0(\sigma_i) P[\sigma_j|\sigma_i]$ (resp. $P_\mathrm{B}[\gamma^*] = p_1(\sigma_j) P[\sigma_i|\sigma_j]$). We have introduced $p_0(\sigma)$ (resp. $p_1(\sigma)$) the probability distribution at the initial time $t_0$ (resp. at the final time $t_1$) of the trajectory. The entropy produced by the trajectory $\gamma$ reads 
\begin{equation} \label{2p}
\Delta_\mathrm{i}S[\gamma] = \log \left( \frac{p_0(\sigma_i)}{p_1(\sigma_j)}\right) +  \log \left( \frac{P[\sigma_j|\sigma_i]}{P[\sigma_i|\sigma_j]}\right).
\end{equation}
The term on the left is called the boundary term, the term on the right the conditional term. 

Let us first characterize the equilibrium distribution $p_\infty (\sigma)$. According to the definition above, it is characterized by $\Delta_\mathrm{i}S[\gamma] = 0$ for all $\gamma$, such that $p_\infty (\sigma_j) P[\sigma_i|\sigma_j] = p_\infty (\sigma_i) P[\sigma_j|\sigma_i]$ for all $(i,j)$. This evidences that the equilibrium distribution fulfills the micro-reversibility condition. Reciprocally, any distribution fulfilling this condition is an equilibrium distribution. 

It is now possible to rewrite Eq.\ref{2p}, 
\begin{equation} \label{2p_bis}
\Delta_\mathrm{i}S[\gamma] = \log \left( \frac{p_0(\sigma_i)}{p_\infty(\sigma_i)}\right) - \log \left( \frac{p_1(\sigma_j)}{p_\infty(\sigma_j)}\right).
\end{equation}
Averaging Eq.\ref{2p_bis} over all forward trajectories yields $\langle \Delta_\mathrm{i}S[\gamma] \rangle_\gamma = -\Delta D_\infty$. $D_\infty$ is the distance to equilibrium, defined as $D_\infty = \Sigma_i p(\sigma_i) (\log p(\sigma_i) - \log p_\infty(\sigma_i) )$. This result evidences that each application of a stochastic map brings the system nearer its equilibrium distribution, which characterizes a relaxation.

At this point it is interesting to consider the textbook case where the stochastic behavior of the system is induced by a thermal bath of temperature $T$.  One denotes $E_k(\sigma)$ the internal energy of the system in the micro-state $\sigma$ at time $t_k$, with $k=0,1$. The system's stochastic internal energy change reads $ \Delta E[\gamma] = E_1(\sigma_j) - E_0(\sigma_i)$. In agreement with the definitions above, $Q[\gamma] = E_0(\sigma_j) - E_0(\sigma_i)$ is the stochastic heat exchanged along $\gamma$, while the stochastic work reads $W[\gamma] =E_1(\sigma_j)-E_0(\sigma_j)$. One gets by construction $W[\gamma] = \Delta E[\gamma] - Q[\gamma]$. 

The probability distribution characterizing thermal equilibrium at $t_0$ is the Boltzmann distribution $p_\infty(\sigma) =  Z^{-1} \exp(- E_0(\sigma)/k_\mathrm{B} T)$. $Z$ is the partition function, that allows us to define the system free energy $F = - \log (Z)$. Let us introduce $\Delta S[\gamma] = -\log(p_0(\sigma_i)) + \log(p_1(\sigma_j))$. $\Delta S[\gamma]$ can be called the system's ``stochastic entropy change": Once averaged over all forward trajectories, it gives back the standard expression for the system's entropy change. Eq.\ref{2p} then becomes $\Delta_\mathrm{i}S[\gamma] = \Delta S[\gamma] - Q[\gamma]/k_\mathrm{B} T$, whose average value is in agreement with the classical definition (Eq. \ref{F}). If the system's initial states of the forward and the backward protocol correspond to the thermal equilibrium (Jarzynski's protocol), one gets $\Delta S[\gamma] = (\Delta E[\gamma]-\Delta F)/k_\mathrm{B} T$. This yields $\Delta_\mathrm{i}S[\gamma] = (W[\gamma]-\Delta F)/k_\mathrm{B} T$, leading to Jarzynski's equality.

\subsection{Generalized integral fluctuation theorem}
A very important achievement of stochastic thermodynamics has been to incorporate information in the expression of a fluctuation theorem, giving rise to the so-called `Generalized Integral Fluctuation Theorem" \cite{Sagawa-Ueda-2012}. A simple intuition can be grasped, again by considering the case of a two-point trajectory that now involves a system and a demon's memory. The system has been read by the demon beforehand, such that the system and the memory state are correlated. Denoting as $x$ and $m$ the system and memory micro-states, $p(x,m)$ (resp. $p(x)$, $p(m)$) the joint (resp. marginal) probabilities, the correlation is quantified by the stochastic mutual information $I(x,m) = \log_2(p(x)p(m)) - \log_2(p(x,m))$. Averaged over the distribution, one recover the usual expression of the mutual information between the system and the demon, $I(S:M) = H_S + H_M - H_{SM}$. 

One can focus on the feedback operation. Namely, the demon exploits its knowledge about the system to perform some operation on it. Supposing that the memory state is not altered by the feedback, the forward (resp. the backward) trajectory reads $\gamma = (x,m,y)$ and $\gamma^* = (y,m,x)$, and their respective probabilities read $P_\mathrm{F}[\gamma] = p_0(x,m) P[y|x,m]$ and $P_\mathrm{B}[\gamma^*] = p_1(y,m) P[x|y,m]$. A feedback is said to be ideal if it is perfectly defined by the memory state, yielding $P[y|x,m] = P[x|y,m]$. We get eventually 

\begin{equation} \label{GIFT}
\Delta_\mathrm{i}S[\gamma] = \log 2 (\Delta S[\gamma] - \Delta I[\gamma]),
\end{equation}
where $\Delta S[\gamma] = \log_2(p_0(x)) - \log_2(p_1(y))$ is the stochastic entropy change of the system. $\Delta I[\gamma] = I_1(y,m) - I_0(x,m)$ is the change of the stochastic mutual information between the system and the memory, where $I_k$ characterizes the joint probability distribution $p_k$. 

Importantly, this expression puts information and entropy production on an equal footing, allowing to quantitatively address the work value of information. The same kind of argument developed for the IFT can indeed be used to demonstrate that $ \Delta S \geq \Delta I(S:M)$, opening a rigorous way to exorcize Maxwell's demon. It basically states that information as quantified by $I(S:M)$ is a resource that can be consumed ($\Delta I(S:M) \leq 0$) to lower the entropy of a system at no work cost. Equivalent expressions can be derived when work is extracted from the protocol, leading to generalized fundamental bounds $W \geq \Delta F + k_\mathrm{B} T \log 2 \Delta I$. This can be used to define efficiencies of Maxwell's demons, e.g. $\eta = W/(\Delta F + k_\mathrm{B} T \log 2 \Delta I)$. Just like usual engines, maximal efficiency is reached when the bound is reached, i.e. when the process is reversibly run.

\subsection{First batch of take-home messages}
Macroscopic thermodynamics has given rise to the concept of engine, as a device that converts heat into work. Maximal efficiency is reached when the device is operated reversibly, connecting the notion of energetic performance to the thermodynamic arrow of time.

These concepts have been extended at the level of single particles by stochastic thermodynamics. Here noise plays a key role to define heat, work, and time arrow. For historical reasons, thermal noise due to the action of thermal baths was first considered. Thus, thermal engines were the first to be designed and experimentally implemented, and they still remain the most studied kind of nano-engines. In the same way, the concepts of entropy production and equilibrium are still widely understood with respect to a thermal bath playing the role of a reference.

However, the framework brought by stochastic thermodynamics is sufficiently general and flexible, such that other kinds of noise can be used as seeds to build ``other" thermodynamical frameworks and explore new physics. We shall adopt this strategy below (See Section \ref{bibi}).

\section{Quantum thermodynamics}

\subsection{Motivations}

Quantum thermodynamics is the converging point of many areas of research, making it a very exciting field where new and transversal concepts are built. Many current lines of research are exposed in reviews and books (See e.g.  \cite{QuThermo}) and it is not my purpose to summarize them here. I shall rather put them in perspective with respect to the legacy of classical thermodynamics presented above, and focus on original research topics developed in my group in the past few years.

Firstly, quantum thermodynamics is the natural follow-up of stochastic thermodynamics, where systems switch from nano to quantum. One of the most important questions is to know what are irreversibility, work and heat in the quantum realm. More technically, one aims to evidence new and genuinely quantum components in fluctuation theorems, that could be related to quantum coherences or entanglement. On the more applied side, one important research line investigates if quantum coherence and correlations can be a resource for nano-engines, that would lead them to outperform their classical counterparts. Reciprocally, what is the energetic cost of fighting against quantum noise?

To answer these questions, a natural scenery is provided by quantum open systems - namely, quantum systems interacting with a driving force and one or several baths. In the equations describing the system's dynamics, the drive's action is usually modeled by some time-dependent Hamiltonian. Hence, the drive can unitarily exchange energy with the system, without changing its von Neumann entropy: This is consistent with the classical definition of work. Conversely, the action of the bath(s) is non-unitary, such that the system's entropy is not necessarily conserved by the interaction: This is reminiscent of a heat exchange. 

However, there is still no consensus on the definitions of heat and work in the quantum realm. One important reason is that the baths interacting with the system are not necessarily at thermal equilibrium. Therefore new thermodynamical concepts must be built, in the absence of temperature. Another reason is genuinely quantum: Quantum measurement disturbs. This is well known, but the energetic consequences of this effect had not been drawn until recently. 

\subsection{Rebuilding quantum thermodynamics on quantum measurement} \label{bibi}

\begin{figure}[t]
        \includegraphics[width=\linewidth]{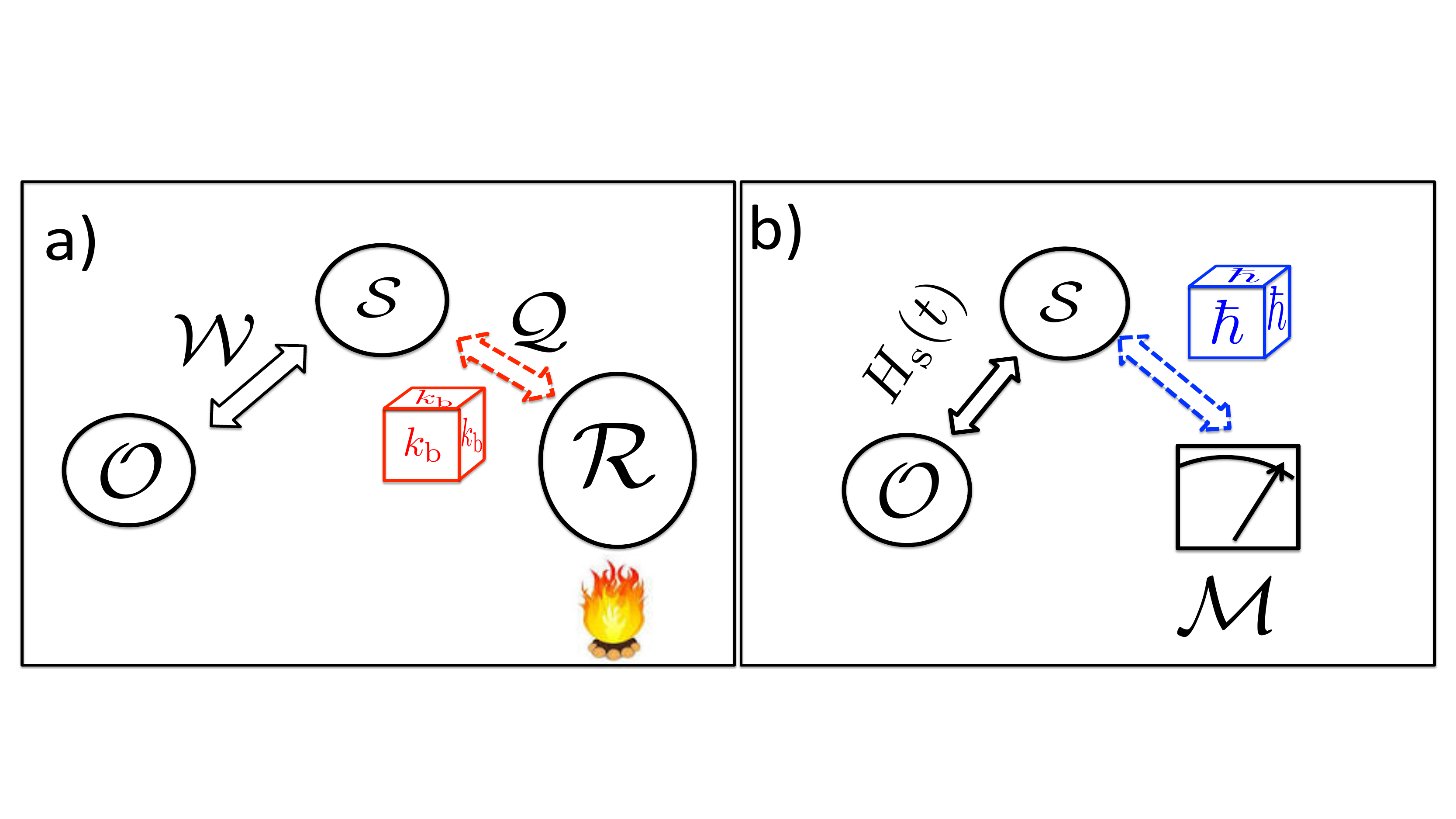}
\caption{Classical versus quantum thermodynamics. A system S exchanges work W with an external controller, and heat Q with a stochastic entity. a) Historical framework. The stochastic entity is a thermal reservoir whose action is symbolized by the dice $k_\mathrm{B}$. b) Rebuilding quantum thermodynamics on quantum measurement. The stochastic entity is a measuring device whose action is symbolized by the dice $\hbar$. }
\label{fig1}
\end{figure}

As mentioned above, a consistent thermodynamical framework can be built for any driven system subjected to noise (See Fig.\ref{fig1}). For historical reasons, the thermal noise was first considered - but there are many other kinds of noise in the quantum world. A fundamental one is the noise induced by projective quantum measurement. The scenery in this case is as simple as it can be: A quantum system evolving under some time-dependent Hamiltonian on the one hand, and projectively measured at discrete times. On the other hand, knowing the outcomes of the measurement and the applied Hamiltonian, it is possible to reconstruct at any time the trajectories of pure quantum states followed by the system, that consists in continuous sequences intertwined by the measurement-induced, stochastic quantum jumps. These quantum trajectories are the quantum counterpart of the stochastic trajectories introduced above, which provided the bread and butter of stochastic thermodynamics. However in the present situation, the stochasticity is genuinely quantum, since it is due to measurement back-action.

For a given quantum trajectory, the system's internal energy along time is identified with the Hamiltonian expectation value along the trajectory. In the spirit of stochastic thermodynamics, work can be defined as the system's energy change during the unitary sequences. ``Heat", on the other hand, can be identified with the sudden energy changes during the quantum jumps. This heat has no classical equivalent, since it comes from the fluctuations induced by measurement back-action. Such fluctuations can only take place, if the measured state has coherences in the basis of the measured observable. It is thus a quantum effect, due to quantum coherences. For this reason, my coworkers and I dubbed it ``quantum heat" \cite{Elouard17Role}. 

Let us now focus on time arrow. For the sake of simplicity, we focus on a protocol defined by some initial eigenstate $\alexket{m_{0}}$ of an observable $\hat{M}$, a unitary evolution $\hat{U}$, and a final measurement of $\hat{M}$ with stochastic result $m_{k_\gamma}$. This elementary quantum trajectory $\gamma$ is perfectly defined by the two points $\gamma = (m_{0}, m_{k_{\gamma}})$. Its probability reads 
$P_\mathrm{F}[\gamma] = P[m_{k_{\gamma}}|m_{0}]$, i.e. $P_\mathrm{F}[\gamma] =  \alexbra{m_{k_\gamma}} \hat{U} \alexket{m_{0}}$. Averaged over all trajectories, the final system state is a mixture $\rho$ of pure states $\alexket{m_k}$ with probabilities $p_{k} = \alexbra{m_{k}} \hat{U} \alexket{m_{0}}$.  
Reciprocally, the bacwkard protocol is defined as follows. Observable $\hat{M}$ is measured on the system, in state $\rho$. Then, the system follows the backward evolution $\hat{U}^\dagger$ after which a final measurement is performed.
The probability of the backward trajectory $\gamma^* = (m_{k_\gamma}, m_{0})$ reads $P_\mathrm{B}[\gamma^*] = p_{k_\gamma} P[m_{0}|m_{k_\gamma}]$. Applying Formula \ref{Si} immediately gives $\Delta_\mathrm{i}S[\gamma] = -\log(p_{k_\gamma})$. Averaged over all trajectories, the entropy production thus reads
\begin{equation}\label{vn}
\langle \Delta_\mathrm{i}S[\gamma] \rangle = \Delta S_\mathrm{VN}
\end{equation}
where $\Delta S_\mathrm{VN}$ is the increase of the Von Neumann entropy of the system along the forward protocol. This expression can easily be extended to multi-points trajectories, or to the case where the protocol does not start with a pure state. Eq.\ref{vn} is extremely important since it connects the Von Neumann entropy - widely used in quantum physics - to entropy production, which is a purely thermodynamical concept. It provides a rigorous demonstration of the well-known ``irreversibility of quantum measurement" that fully exploits the relevant framework of stochastic thermodynamics. In particular, it now allows measuring ``how much irreversibility'' a measurement creates.

\subsection{Measurement driven engines}

\begin{figure}[t]
        \includegraphics[width=\linewidth]{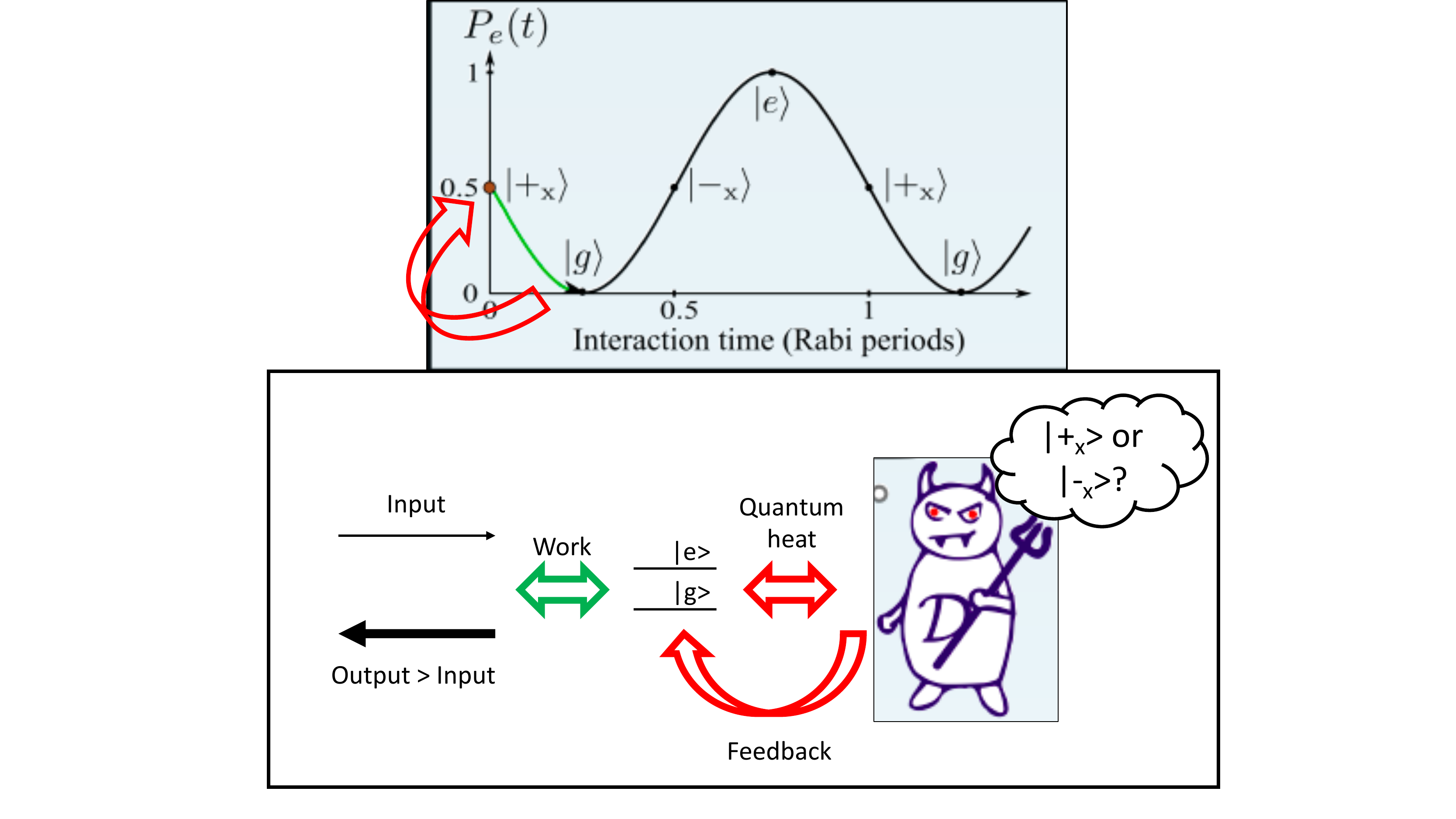}
\caption{Top. Population of the driven qubit excited state $P_e(t)$ (Rabi oscillation). The state $\alexket{+_x}$ (resp. the state $\alexket{-_x}$) gives rise to maximal work extraction in the field (resp. from the field). Bottom. Scheme of a quantum Maxwell's demon experiment. The qubit exchanges work with the driving field, while being periodically measured by the demon in the $\{ \alexket{+_x}; \alexket{-_x} \}$ basis. If the demon measures $\alexket{-_x}$, it performs a feedback on the qubit to bring it in the state $\alexket{+_x}$, which is favorable for work extraction.}
\label{fig2}
\end{figure}

It should now be clear that projective measurement, just like any stochastic process, can be seen as a source of irreversibility and energy - playing a role quite similar to the good old thermal bath. In particular, a non-zero amount of quantum heat can be exchanged {\it on average} between the system and the measurement channel, as soon as the measured observable does not commute with the system's Hamiltonian. Building on this analogy, my coworkers and I have suggested to use quantum measurement as a new kind of energetic resource that could fuel quantum engines (See \cite{Elouard17Extracting} and Fig.\ref{fig2}). The experiment that we have suggested involves a qubit of energy eigenstates denoted $\alexket{0}$ and $\alexket{1}$, of transition frequency $\omega_0$ as a working substance, that exchanges work with some resonant driving field. The mechanism is a classical Rabi oscillation, where the qubit's state evolves as $\alexket{\psi(t)} = \cos(\Omega t/2) \alexket{0} + \sin(\Omega t/2) \alexket{1}$ with $\Omega$ the classical Rabi frequency. Work extraction takes place during stimulated emission, when the qubit provides energy to the field. Maximal power extraction is reached when the qubit is in the coherent superposition $\alexket{+} = ( \alexket{0} +  \alexket{1} )/\sqrt{2}$ which gives rise to the maximal slope for the Rabi oscillation. 

To cyclically extract work, the strategy we suggested was to use a measurement followed by a feedback loop, to stabilize the qubit in the state $\alexket{+}$. One cycle consists of the four following steps: (i) Work extraction: after the qubit is initialized in the state $\alexket{+}$, it evolves in the state $\alexket{\phi(\tau)} = \cos(\Omega \tau/2) \alexket{+} + \sin(\Omega \tau/2) \alexket{-}$  while an amount of work $W = \hbar \omega_0 \sin(\Omega \tau/2)$ is extracted by the field (ii) Measurement of the qubit in the $\{  \alexket{+} ;  \alexket{-}  \}$ basis where $\alexket{-} = ( -\alexket{0} +  \alexket{1} )/\sqrt{2}$. During this step, an amount of quantum heat $Q_\mathrm{q} = W$ is provided by the measurement channel to the qubit (iii) Feedback to prepare the state $ \alexket{+} $. This feedback step costs no energy since the states $\alexket{+}$ and $\alexket{-}$ have the same energy (iv) Erasure of the classical memory used to control the feedback loop. The energetic cost of this process is lower bounded by Landauer's erasure work $W_L = k_\mathrm{B} T H[p]$ where $T$ is the temperature of the memory and $H[p]$ its Shannon's entropy. 

Actually, this machine is nothing but a new kind of Maxwell's demon engine. The feature that makes it really quantum is that it does not extract energy from a hot thermal bath, but from the measurement process itself. Thus the two facets of quantum measurement are exploited in this device: Measurement not only allows to extract information, but it also provides energy since it back-acts on the system's state. Stated in fancy words, with such an engine you can put a body in motion,`` just by looking at it". 

For an engine, an important figure of merit is its yield $\eta$. It is computed, by comparing the net extracted work $W-W_L$ to the consumed resource $Q_\mathrm{q}$ i.e. $\eta = 1 - W_L/Q_\mathrm{q}$. Interestingly, $\eta \rightarrow 1$ when $\Omega \tau \ll 1$.  This corresponds to the Zeno regime, where measurements are performed at such a fast rate that the qubit is ``frozen" in the $\alexket{+}$ state. In this situation, the measurement outcome is certain and the memory's entropy vanishes, such that no erasure is needed. Reaching such a yield means that the quantum heat provided by the measurement channel is fully converted into work, the engine behaving as a transducer. The power is another relevant figure of merit. It turns out that maximal power is also reached in the Zeno regime, as a simple consequence of the fact that $\alexket{+}$ is the best state for power extraction - as mentioned above. Unlike classical engines where one has to chose between maximal efficiency and maximal power, the present device allows to operate at maximal efficiency and power simultaneously. This is typical of the fact that we have now departed from standard thermodynamics and that new intuitions must be built. 

\subsubsection*{Discussion}
The engine presented above was the first to explicitly exploit ``quantum heat", i.e. measurement induced back-action, for the sake of work extraction. Obviously, the main value of the proposal is not its practical interest. It is a proof of concept, that evidences the reality of energy exchanges with a measurement channel. Since then, the concept of quantum heat has bloomed to give rise to new proposals for measurement powered engines \cite{Elouard18Engine,Talkner18,JEA19}, to cool down qubits \cite{Campisi19}, or to track entanglement generation \cite{EAH19}. Let us stress that the engines proposed in refs. \cite{Campisi19} and \cite{Talkner18} need no feedback control to function, clearly showing that quantum heat contains more than information. Since decoherence is nothing but an unwanted measurement performed by some uncontrolled environment, quantum heat is also expected to be a relevant concept to estimate the energetic costs related to feedback-based stabilization \cite{Elouard17Role}. 

The quantum heat is the energetic counterpart of the measurement postulate and as such, it can be perceived differently by different users of quantum theory. It can be seen as a practical, effective quantity allowing to take quantitatively into account the effect of a measurement on thermodynamical quantities. An interesting line of research now will consist in ``opening the black box", i.e. modeling the measurement process itself and track the energetic and entropic fluxes within the measurement channel. Just like in classical thermodynamics, where heat and irreversibility are expected to vanish when one reaches a complete, ultimate description of the system, one could thus expect to find that quantum heat is not a fundamental concept. 

On the other hand, one can be of the opinion that quantum mechanics entirely relies on some act of measurement, and that whichever the degree of precision in the described mechanism, its very meaning is built on the irreversible records of measurement outcomes. In this view, the quantum heat rather appears as a fundamental concept that will always be part of the thermodynamical description. Current debates about heat and work in quantum thermodynamics reflect in some sense the still ongoing debates about the status of the measurement postulate - quantum thermodynamics providing a new playground, and -why not?- Maybe new ideas and new experiments to explore interpretations of quantum mechanics \cite{GA18, JEA19}.

\subsection{Thermodynamics of quantum computing}
As a final promising field of investigations for quantum thermodynamics, it is worth mentioning the thermodynamical study of quantum computing in the context of the present school. As a matter of fact, the ``quantum advantage" usually put forward to motivate quantum versus classical computing is its reduction in complexity. However, another advantage is that quantum computing is in principle reversible. To be more specific, an ideal quantum algorithm like the Deutsch problem consists in initializing the data register in the state $\alexket{0}$ of the computational basis, a unitary operation, and a final measurement in the computational basis. In the absence of noise, the algorithm can be seen as a quantum interference: The register quantum state before the final measurement pertains to the computational basis and the outcome provides a noise-free answer to the asked question. In particular, there is no back-action associated to this final measurement, i.e. the act of measurement has no effect on the system's state. Therefore, it is conceivable that the agent performing the computation can record the outcome, and then reverse the whole protocol to bring back the register into its initial state $\alexket{0}$, such that there is no heat dissipation associated to the reset. \\

Conversely, activating the gates of the quantum circuit is not energetically free. Simply considering a single qubit gate, the minimal energetic cost to run it is reached when we let the qubit resonantly interact with a coherent field during a well-defined time. The energetic cost associated to this operation can be quantified as the minimal number of photons put in the coherent field. 
To reduce the cost, it is tempting to work with a small number of photons. However, small fields get entangled with the qubit, leading to some fundamental noise affecting the process as soon as the field is traced out \cite{Miko17,Banacloche02}. Preliminary studies show that at least $1000$ photons are needed in the field to keep a sufficiently good fidelity on the gate. For microwave photons interacting with superconducting qubits, the energetic bill to activate a single gate is typically $10^{-21}$J. This is the same order of magnitude as the ultimate heat dissipated by the erasure of a single bit. Large scale quantum computers will involve a large number of gates, in particular for error correction. As a consequence, energy will quite certainly play a key role to benchmark future quantum computing architectures.

\section*{Acknowledgements}
It is a pleasure to thank Ioan Pop, Benjamin Huard and Michel Devoret for their kind invitation to provide this manuscript, as well as Patrice Camati and Michele Campisi for their careful reading.

% TODO: include funding information
\paragraph{Funding information}
This work was supported by the Foundational Questions Institute Fund (Grant number FQXi-IAF19-05 and FQXi-IAF19-01), the Templeton World Charity Foundation, Inc (Grant No. TWCF0338) and the ANR Research Collaborative Project ``Qu-DICE" (ANR-PRC-CES47).

\nolinenumbers

\end{document}